%% file: main.tex
\documentclass[10pt,conference]{IEEEtran}
\IEEEoverridecommandlockouts
\usepackage{cite}
\usepackage{amsmath,amssymb,amsfonts}
\usepackage{algorithmic}
\usepackage{graphicx}
\usepackage{textcomp}
\usepackage{xcolor}
\usepackage{hyperref}
\usepackage[hyphenbreaks]{breakurl}

\usepackage{multirow}
\def\BibTeX{{\rm B\kern-.05em{\sc i\kern-.025em b}\kern-.08em
    T\kern-.1667em\lower.7ex\hbox{E}\kern-.125emX}}
\begin{document}

\title{Carbon Emissions of Quantum Circuit Simulation: More than You Would Think}
\author{\IEEEauthorblockN{
Jinyang Li\textsuperscript{\dag},
Qiang Guan\textsuperscript{\S}, 
Dingwen Tao\textsuperscript{\ddag},
Weiwen Jiang\textsuperscript{\dag}}

\IEEEauthorblockA{\textsuperscript{\dag}George Mason University, Department of Electrical and Computer Engineering, VA, USA.\\
\textsuperscript{\S}Kent State University, OH, USA.\\
\textsuperscript{\ddag}Indiana University, IN, USA.\\
\{jli56, wjiang8\}@gmu.edu
\vspace{-0.15in}}
}
\maketitle

\begin{abstract}
The rapid advancement of quantum hardware brings a host of research opportunities and the potential for quantum advantages across numerous fields. In this landscape, quantum circuit simulations serve as an indispensable tool by emulating quantum behavior on classical computers. They offer easy access, noise-free environments, and real-time observation of quantum states. However, the sustainability aspect of quantum circuit simulation is yet to be explored. In this paper, we introduce for the first time the concept of environmental impact from quantum circuit simulation. We present a preliminary model to compute the CO$_2$e emissions derived from quantum circuit simulations. Our results indicate that large quantum circuit simulations (43 qubits) could lead to CO$_2$e emissions 48 times greater than training a transformer machine learning model.
\end{abstract}


\input{introduction}
\input{model}

\input{result}

\input{conclusion}
\bibliographystyle{IEEEtran}
\bibliography{jinyang}

\end{document}

%% file: introduction.tex
\section{introduction}
Quantum computing, recognized as the next technological revolution, holds immense potential to transform a wide range of areas, including cryptography, materials science, and AI. Currently, however, quantum hardware is both limited and expensive to access, making its widespread adoption challenging. While quantum circuit simulation has emerged as a supporting tool. It employs classical computing resources to emulate the behavior of quantum circuits, thereby bypassing the need for physical quantum hardware. Quantum circuits comprise quantum bits (qubits) and quantum gates, which manipulate these qubits. 
For example, State Vector Simulators, simulate a quantum circuit by computing the wavefunction of the qubit’s statevector as gates and instructions are applied. 
There are also cloud-based simulators from most of the major quantum cloud providers such as IBM.

The importance of quantum circuit simulation can be attributed to several reasons: \textbf{\textit{Limited Quantum Hardware Availability:}} Quantum computing resources are scarce, expensive, and often accessible only through cloud-based platforms with waiting times. And this situation could be worse as the increased demand for cloud resources\cite{ravi2022quantum}. \textbf{\textit{Noise Influence:}} In the "Noisy Intermediate-Scale Quantum"(NISQ) era\cite{Preskill2018quantumcomputingin}, where quantum computers have limited qubits and high error rates, quantum circuit simulators play a vital role. They enable researchers to develop, test, and optimize quantum algorithms in an ideal, noise-free environment before deploying them on physical quantum hardware. \textbf{\textit{Algorithm Testing and Development:}} Quantum circuit simulators, unlike actual quantum systems, allow real-time observation of a computation's state. Measuring states in quantum systems is challenging due to the disruption of computation and partial view of the quantum state. Simulators, however, mimic quantum states on classical computers, enabling developers to inspect the full quantum state at any moment, facilitating the development of algorithms such as Variational Quantum Circuit(VQC)\cite{jiang2021co, wang2021exploration, peddireddy2022classical,hu2022design, hu2022quantum, wang2023qumos, liang2021can}, and deepening quantum programming understanding. \textbf{\textit{Quantum Error Mitigation:}} Quantum systems are inherently noisy, and quantum error correction is still in its infancy. Simulations can help model the noise characteristics of quantum devices and develop error mitigation techniques\cite{liu2022classical}.

Conventionally, the evaluation metrics for quantum circuit simulations include simulation fidelity, computational speed, and resource usage. However, as with any computational process, quantum circuit simulation comes with an energy cost, which translates into substantial CO$_2$e emissions. Considering the global need to reduce greenhouse gas emissions, understanding the carbon footprint of quantum circuit simulations is as crucial as enhancing their performance. A balance needs to be struck between the pursuit of technological advancement and environmental sustainability. This sphere of interest encompasses diverse stakeholders: researchers and developers focused on quantum algorithm design, hardware companies like NVIDIA who are advancing quantum simulation projects such as cuQuantum\cite{cuquantum}, and organizations advocating for sustainable practices due to the environmental implications.
\begin{table}[t]
\caption{comparison of co$_2e$ emissions from human activities, and classical machine learning training.}
\label{tab: table1}
\tabcolsep 17 pt
\renewcommand{\arraystretch}{1.1}
\begin{tabular}{c|c}
\hline
Activity                            & CO$_2$e (kg) \\ \hline
Household electricity use for a day \cite{life3} & 12.44           \\
Driving a car for 100 km \cite{life1}           & 24.80              \\
Transatlantic flight, 1 passenger \cite{life2}  & 313.90              \\ 
\hline
Transformer\_base training  \cite{strubell2019energy}                  & 11.79              \\ 
Transformer\_large training  \cite{strubell2019energy}                 & 87.09              \\ 
ELMo training  \cite{strubell2019energy}                             & 118.84             \\ \hline
Quantum Circuit Simulation (43 qubits)                               & 568.77             \\ \hline

\end{tabular}
\end{table}
To further illustrate the environmental impact of quantum circuit simulations, we compare their energy consumption and emissions with those of common life activities and classical machine training, as shown in Table \ref{tab: table1}. The result reveals that the CO$_2$e emissions resulting from quantum circuit simulations can exceed those of other activities and processes. For instance, the simulation can generate up to 1.81 times the CO$_2$e emissions of a one-way flight from New York to London. Notably, when compared with classical computing, the quantum circuit simulation demonstrates an even more pronounced environmental impact; it produces approximately 48 times the CO$_2$e emissions of training a standard transformer base model.
The main contributions of this paper are: \textbf{(1)} Bring the notion of environmental impact from quantum circuit simulation. \textbf{(2)} Build the initial model for calculating the CO$_2$e emissions of simulation.

%% file: model.tex
\section{energy consumption and carbon footprint of quantum circuit simulation}

The carbon emissions from quantum circuit simulations derive from a multitude of sources. This includes Embodied Emissions, encompassing the carbon footprint from the manufacturing and disposal of hardware, Idle Power Consumption, representing the emissions when the system is powered but not actively processing, and Dynamic Power Consumption, which relates to active processing and data transfer. Dynamic Power Consumption is affected by the properties of a given quantum circuit, such as the number of qubits, the circuit depth, etc.
Besides, it also hinges on other factors, such as the computational resources utilized, their efficiency, and the simulation duration. These resources include processors (CPU/GPU), memory modules, cooling systems, and a multitude of peripheral devices. 
Note that this investigation primarily focuses on Dynamic Power Consumption for quantum circuit simulations, but the proposed model can be easily extended to support other factors such as the load on the processors, the utilization of memory, and the efficiency of the cooling systems.


To formulate the simulation-emission model, we first define the system to run the simulation as follows.
For a granular estimate of the energy consumed, we factor in the number of processors (`n'), the average power per processor (`Pp' in kW), and the simulation duration (`T' in hours). 
The environmental impact, measured as Carbon Dioxide Equivalent (CO$_2$e) emissions, is then determined by the Carbon Intensity (`CI'), a measure of CO$_2$e emissions per unit of electricity consumed (in kg/kWh), and the Power Usage Effectiveness (`PUE'). The PUE, a measure of data center efficiency, describes the proportion of total power consumption utilized directly by the computing equipment.

Incorporating the above definitions, the CO$_2$e emissions are calculated as: $CO_2e = n \times Pp \times T \times CI \times PUE$. This comprehensive approach allows us to estimate the precise environmental implications of quantum circuit simulations.
For example, consider a quantum circuit simulation on a personal computer with an average power draw of $P = 0.04276$ kW over an execution time of $T = 0.01861$ hours. The energy consumption for this simulation would be $E = 0.04276\, \text{kW} \times 0.01861\, \text{hours} = 0.00080\, \text{kWh}$.
Furthermore, using a Carbon Intensity value of $CI = 0.429$ kg/kWh, as per the average datacenter carbon emissions~\cite{EIA}, and a Power Usage Effectiveness of $PUE = 1.58$, reflecting the average industry datacenter PUE~\cite{patterson2021carbon}, the CO$_2$e emissions can be computed as $CO_2e = 0.00080\, \text{kWh} \times 0.429\, \text{kg/kWh} \times 1.58 = 0.00054\, \text{kg}$.








%% file: result.tex
\section{results}

\begin{figure}[t]
    \centering
    \includegraphics[width=1\columnwidth]{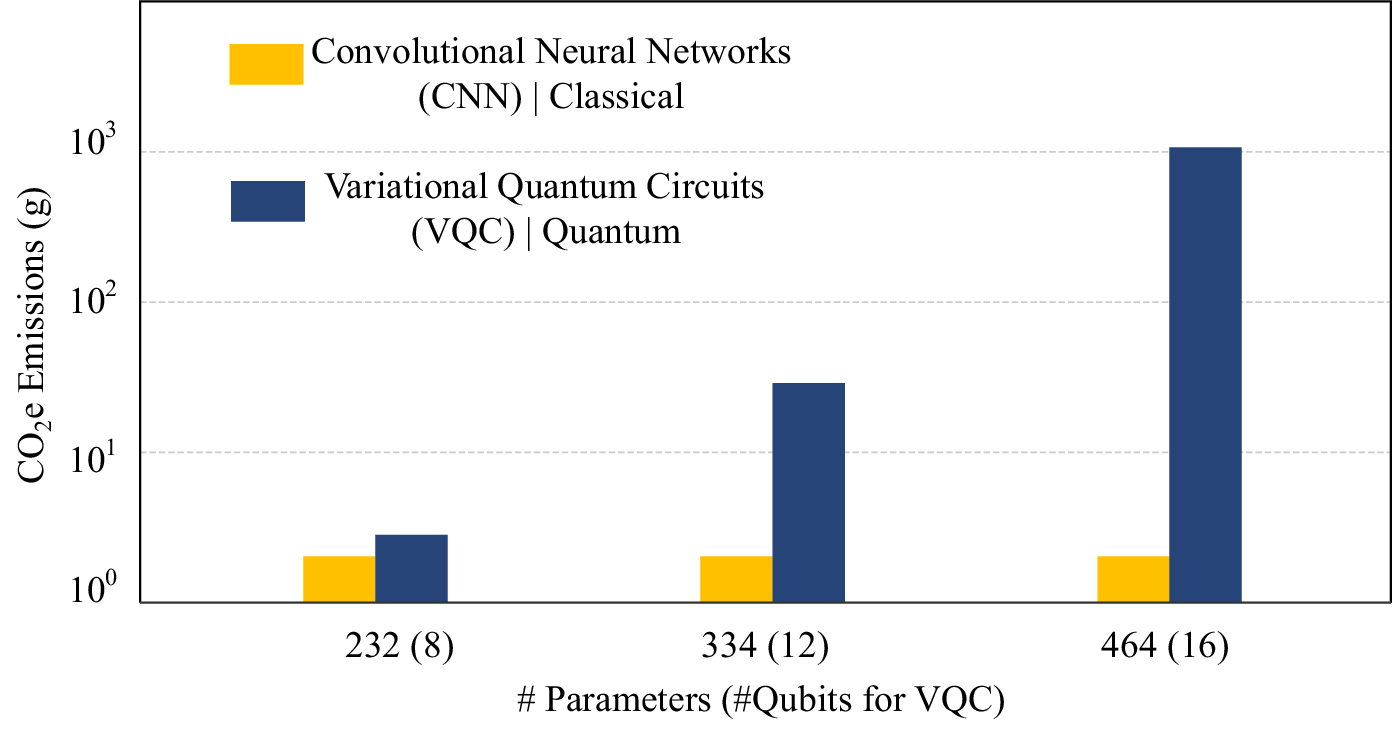}
    \vspace{-20 pt}
    \caption{CO$_2$e Emission Result for CNN and VQC.}
    \label{fig: figure1}
\end{figure}

\subsection{Small Quantum Circuit Simulations}
We first conducted our own simulations on small quantum circuits to examine the impact of certain parameters on energy consumption and emissions. However, due to that a single run of quantum circuit simulation was rather too fast, which would make less sense in the context of analyzing the energy cost and emission, we set the experiments on the task of quantum machine learning. That is a very promising field that takes the power of quantum computing in machine learning tasks. In our experiment, we used the MNIST dataset for training purposes, comprising 20,000 training samples. The training was set for 20 epochs with a batch size of 256. We tracked the execution time during these runs, alongside the average power consumption for each processor core, to calculate the final CO$_2$e emission. The results of these measurements are depicted in Figure\ref{fig: figure1}.
In addition, we conducted parallel experiments with classical neural network models, ensuring that the number of parameters corresponded with those in the quantum circuits. The outcomes from these classical models are represented by the red bars in Figure \ref{fig: figure1}.

\begin{table}[t]
\caption{emission result from aws large-scale quantum circuit simulation.}
\label{tab: table2}
\begin{tabular}{|c|c|c|c|}
\hline
\#Qubit & \#Execution time (hours) & \#Processor & CO$_2$e emissions (kg) \\ \hline
40      & 1.25                     & 256         & 54.23              \\ \hline
41      & 1.41                     & 512         & 122.91             \\ \hline
42      & 1.58                     & 1024        & 274.74             \\ \hline
43      & 1.64                     & 2048        & 568.77             \\ \hline
\end{tabular}
\vspace{-15pt}
\end{table}

\subsection{Large Quantum Circuit Simulations}
In this section, we analyzed existing data on large quantum circuit simulations. Due to the limitation of the device, we can't now scale up the quantum circuit simulation, however, we can still collect the experiment results from large companies such as NVIDIA or AWS. Here is the emission result based on the classic simulations for quantum circuits from AWS \cite{aws}. To be noticed, the result here in Table \ref{tab: table2} is a single run of simulation instead of a training VQC. It is evident that the CO$_2$e emissions escalate as the number of qubits increases, meriting attention and concern during the development of quantum circuit simulation.



%% file: conclusion.tex
\section{conclusion}
By analyzing the energy consumption and carbon emissions, we can better comprehend the environmental footprint of quantum circuit simulations, thus enabling us to devise strategies to reduce their impact.

%% file: main.bbl
\begin{thebibliography}{10}
\providecommand{\url}[1]{#1}
\csname url@samestyle\endcsname
\providecommand{\newblock}{\relax}
\providecommand{\bibinfo}[2]{#2}
\providecommand{\BIBentrySTDinterwordspacing}{\spaceskip=0pt\relax}
\providecommand{\BIBentryALTinterwordstretchfactor}{4}
\providecommand{\BIBentryALTinterwordspacing}{\spaceskip=\fontdimen2\font plus
\BIBentryALTinterwordstretchfactor\fontdimen3\font minus
  \fontdimen4\font\relax}
\providecommand{\BIBforeignlanguage}[2]{{%
\expandafter\ifx\csname l@#1\endcsname\relax
\typeout{** WARNING: IEEEtran.bst: No hyphenation pattern has been}%
\typeout{** loaded for the language `#1'. Using the pattern for}%
\typeout{** the default language instead.}%
\else
\language=\csname l@#1\endcsname
\fi
#2}}
\providecommand{\BIBdecl}{\relax}
\BIBdecl

\bibitem{ravi2022quantum}
G.~S. Ravi, K.~N. Smith, P.~Gokhale, and F.~T. Chong, ``Quantum computing in
  the cloud: Analyzing job and machine characteristics,'' 2022.

\bibitem{Preskill2018quantumcomputingin}
\BIBentryALTinterwordspacing
J.~Preskill, ``Quantum {C}omputing in the {NISQ} era and beyond,''
  \emph{{Quantum}}, vol.~2, p.~79, Aug. 2018. [Online]. Available:
  \url{https://doi.org/10.22331/q-2018-08-06-79}
\BIBentrySTDinterwordspacing

\bibitem{jiang2021co}
W.~Jiang, J.~Xiong, and Y.~Shi, ``A co-design framework of neural networks and
  quantum circuits towards quantum advantage,'' \emph{Nature communications},
  vol.~12, no.~1, pp. 1--13, 2021.

\bibitem{wang2021exploration}
Z.~Wang, Z.~Liang, S.~Zhou, C.~Ding, Y.~Shi, and W.~Jiang, ``Exploration of
  quantum neural architecture by mixing quantum neuron designs,'' 2021.

\bibitem{peddireddy2022classical}
D.~Peddireddy, V.~Bansal, and V.~Aggarwal, ``Classical simulation of
  variational quantum classifiers using tensor rings,'' 2022.

\bibitem{hu2022design}
Z.~Hu, J.~Li, Z.~Pan, S.~Zhou, L.~Yang, C.~Ding, O.~Khan, T.~Geng, and
  W.~Jiang, ``On the design of quantum graph convolutional neural network in
  the nisq-era and beyond,'' in \emph{2022 IEEE 40th International Conference
  on Computer Design (ICCD)}.\hskip 1em plus 0.5em minus 0.4em\relax IEEE,
  2022, pp. 290--297.

\bibitem{hu2022quantum}
Z.~Hu, P.~Dong, Z.~Wang, Y.~Lin, Y.~Wang, and W.~Jiang, ``Quantum neural
  network compression,'' in \emph{Proceedings of the 41st IEEE/ACM
  International Conference on Computer-Aided Design}, 2022, pp. 1--9.

\bibitem{wang2023qumos}
Z.~Wang, J.~Li, Z.~Hu, B.~Gage, E.~Iwasawa, and W.~Jiang, ``Qumos: A framework
  for preserving security of quantum machine learning model,'' \emph{arXiv
  preprint arXiv:2304.11511}, 2023.

\bibitem{liang2021can}
Z.~Liang, Z.~Wang, J.~Yang, L.~Yang, Y.~Shi, and W.~Jiang, ``Can noise on
  qubits be learned in quantum neural network? a case study on quantumflow,''
  in \emph{2021 IEEE/ACM International Conference On Computer Aided Design
  (ICCAD)}.\hskip 1em plus 0.5em minus 0.4em\relax IEEE, 2021, pp. 1--7.

\bibitem{liu2022classical}
J.~Liu, A.~Gonzales, and Z.~H. Saleem, ``Classical simulators as quantum error
  mitigators via circuit cutting,'' 2022.

\bibitem{cuquantum}
\BIBentryALTinterwordspacing
T.~Lubowe and S.~Morino, ``Best-in-class quantum circuit simulation at scale
  with nvidia cuquantum appliance,'' 2022. [Online]. Available:
  \url{https://developer.nvidia.com/blog/best-in-class-quantum-circuit-simulation-at-scale-with-nvidia-cuquantum-appliance/}
\BIBentrySTDinterwordspacing

\bibitem{life3}
\BIBentryALTinterwordspacing
``Frequently asked questions (faqs) - how much electricity does an american
  home use?'' U.S. Energy Information Administration (EIA). [Online].
  Available: \url{https://www.eia.gov/tools/faqs/faq.php?id=97\&t=3}
\BIBentrySTDinterwordspacing

\bibitem{life1}
\BIBentryALTinterwordspacing
``Greenhouse gas emissions from a typical passenger vehicle,'' U.S.
  Environmental Protection Agency. [Online]. Available:
  \url{https://www.epa.gov/greenvehicles/tailpipe-greenhouse-gas-emissions-typical-passenger-vehicle}
\BIBentrySTDinterwordspacing

\bibitem{life2}
\BIBentryALTinterwordspacing
``Carbon emissions calculator,'' International Civil Aviation Organization
  (ICAO). [Online]. Available:
  \url{https://applications.icao.int/icec/Home/Index}
\BIBentrySTDinterwordspacing

\bibitem{strubell2019energy}
E.~Strubell, A.~Ganesh, and A.~McCallum, ``Energy and policy considerations for
  deep learning in nlp,'' 2019.

\bibitem{EIA}
\BIBentryALTinterwordspacing
``How much carbon dioxide is produced per kilowatt hour of u.s. electricity
  generation?'' U.S. Energy Information Administration (EIA), 2021. [Online].
  Available: \url{https://www.eia.gov/tools/faqs/faq.php?id=74\&t=11}
\BIBentrySTDinterwordspacing

\bibitem{patterson2021carbon}
D.~Patterson, J.~Gonzalez, Q.~Le, C.~Liang, L.-M. Munguia, D.~Rothchild, D.~So,
  M.~Texier, and J.~Dean, ``Carbon emissions and large neural network
  training,'' 2021.

\bibitem{aws}
\BIBentryALTinterwordspacing
F.~Baruffa and P.~Lougovski, ``Simulating 44-qubit quantum circuits using aws
  parallelcluster.'' [Online]. Available:
  \url{https://aws.amazon.com/cn/blogs/hpc/simulating-44-qubit-quantum-circuits-using-aws-parallelcluster/}
\BIBentrySTDinterwordspacing

\end{thebibliography}
